\documentclass[%
 reprint,
 amsmath,amssymb,
 aps,
 nofootinbib,
 superscriptaddress
]{revtex4-2}

\usepackage{graphicx}
\usepackage{dcolumn}
\usepackage{bm}
\RequirePackage{lmodern}      
\RequirePackage{amssymb}
\RequirePackage{amsfonts}
\RequirePackage{amsmath}

\usepackage[usenames,dvipsnames]{xcolor}

\definecolor{darkCyan}{RGB}{0, 139, 139}
\definecolor{darkMagenta}{RGB}{139, 0, 139}

\usepackage{hyperref}

\usepackage{mathtools}

\usepackage{tikz}
\usepackage{tikz-cd}
\usetikzlibrary{decorations.markings,arrows.meta }
\usetikzlibrary{arrows,shapes.geometric,positioning}
\usepackage{adjustbox}

\usepackage{subcaption}
\usepackage{extarrows}
\usepackage{bbm}
\usepackage{mathtools,slashed}
\usepackage{tablefootnote}
\usepackage{diagbox}
\usepackage{multirow}
\usepackage{graphbox}
\usepackage{fnpct}
\usepackage{dsfont}
\usepackage{stmaryrd} 
\usepackage{relsize}
\usepackage{orcidlink}

\usepackage[most]{tcolorbox}
\tcbset{
    enhanced jigsaw, 
    colback=gray!5!white,
    colframe=gray!20!white,
    coltitle=black,
    fonttitle=\bfseries
}
\newtcolorbox[auto counter,number within=section]{mybox}[2][]{%
fonttitle=\bfseries,
title=Box \thetcbcounter: #2,#1
}

\tikzstyle{basic} = [rectangle, text centered, fill=gray!10]

\newcommand{\coactionurl}{https://frwcoaction.ca/}

\newcommand{\coactionsymbolAlt}{%
     \includegraphics[align=c, scale=1.5]{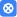}
}

\newcommand{\webappAlt}{\href{\coactionurl}{\coactionsymbolAlt}}

\usepackage{fontawesome5}



\renewcommand{\d}{\mathrm{d}}
\newcommand{\res}{\text{Res}}
\newcommand{\disc}{\text{Disc}}
\newcommand{\kin}{\text{kin}}
\newcommand{\vep}{\varepsilon}
\newcommand{\vphi}{\varphi}

\newcommand{\ra}{\rangle}
\newcommand{\la}{\langle}
\newcommand{\dlog}{\mathrm{dlog}}

\newcommand{\G}{\mathcal{G}}
\newcommand{\g}{{\mathfrak{g}}}
\newcommand{\h}{{\mathfrak{h}}}

\newcommand{\V}{\mathcal{V}}
\newcommand{\B}{\mathcal{B}}
\newcommand{\Csf}{{\mathsf{C}}}
\newcommand{\csf}{\mathsf{c}}

\newcommand{\Rsf}{\mathsf{R}}
\newcommand{\rsf}{\mathsf{r}}
\newcommand{\nn}{\nonumber}

\newcommand{\mbf}[1]{\mathbf{#1}}

\newcommand{\be}{\begin{equation}\begin{aligned}}
\newcommand{\ee}{\end{aligned}\end{equation}}

\DeclareMathOperator{\sgn}{sgn}

\newcommand{\solidHalfEdge}[1]{
        \begin{tikzpicture}[scale=#1]
            \coordinate (A) at (0,0);
            \coordinate (B) at (1/4,0);
            \draw[double,gray,thick] (A) -- (B);
            \fill[black] (A) circle (1.8pt);
        \end{tikzpicture}
}

\newcommand{\twoChaing}{\includegraphics[scale=.35]{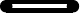}}
\newcommand{\twoChaingg}{\includegraphics[scale=.35]{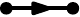}}
\newcommand{\twoChainggg}{\includegraphics[scale=.35]{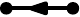}}
\newcommand{\twoChaingggg}{\includegraphics[scale=.35]{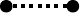}}

\newcommand{\al}{\tikz \filldraw[scale=0.1, rotate=90] (0,0) -- (1,0) -- (0.5,0.866) -- cycle;}
\newcommand{\ar}{\tikz \filldraw[scale=0.1, rotate=-90] (0,0) -- (1,0) -- (0.5,0.866) -- cycle;}

\begin{document}


\title{A Graphical Coaction for FRW Wavefunction Coefficients}

\author{Andrew J.~McLeod\orcidlink{0000-0001-7685-2929}}
\email{andrew.mcleod@ed.ac.uk}
\affiliation{%
    Higgs Centre for Theoretical Physics,
    School of Physics and Astronomy,
    The University of Edinburgh,
    Edinburgh EH9 3FD, Scotland, UK
}

\author{Andrzej Pokraka\orcidlink{0000-0003-1186-4624}}
\email{a.m.pokraka@uva.nl}
\affiliation{%
        Institute of Theoretical Physics,
        University of Amsterdam,
        Amsterdam, 1098 XH, The Netherlands
}

\author{Lecheng Ren\orcidlink{0000-0002-0846-8017}}
\email{lecheng.ren@qmul.ac.uk}
\affiliation{%
    Centre for Theoretical Physics,
    Department of Physics and Astronomy,
    Queen Mary University of London, E1 4NS, UK
}

\begin{abstract}%
We show that the wavefunction of the universe in theories of conformally coupled scalars in power-law Friedmann-Robertson-Walker (FRW) cosmologies satisfies a graphical coaction, by means of which we can understand its complete analytic structure in terms of the acyclic minors of Feynman graphs.
Our construction extends to all particle multiplicities and any loop order, and if we isolate certain weight-one contributions, it reproduces the ``kinematic flow'' that encodes the differential equation of the wavefunction coefficients. 
In a similar manner, the (sequential) discontinuities of wavefunction coefficients can also be extracted from the coaction.
The coaction for any graph has been implemented in a user-friendly web application at \webappAlt\!\!. 
\end{abstract}

\maketitle
\allowdisplaybreaks

\section{Introduction}


Conformally-coupled scalar field theories in FRW cosmologies provide enlightening toy models with a rich mathematical structure and a connection to inflationary physics. 
These theories are usually studied in the presence of (non-conformal) polynomial interactions where 
$$\d s^2 = a^2(\eta) \left[-\d \eta^2 + {\textstyle \sum_{i=1}^d} \d x_i \d x_i \right]\,,$$
is the FRW metric,  $\eta \in (-\infty,0]$ is the conformal time, and the scale factor is a power-law $a(\eta) = (\eta/\eta_0)^{-(1+\varepsilon)}$ characterized by the cosmological parameter $\varepsilon$. 
This scale factor specializes to several well-studied cosmological scenarios including de Sitter ($\varepsilon = 0$), flat ($\varepsilon = -1$), radiation-dominated ($\varepsilon = -2$), and matter-dominated ($\varepsilon = -3$) universes~\cite{De:2023xue, Arkani-Hamed:2023kig}. Moreover, for $0<\varepsilon\ll1$, this model is expected to capture the essential dynamics of  near de Sitter inflationary cosmology.

In de Sitter ($\vep{=}0$) theories, equal-time correlation functions encode the quantum fluctuations at the end of inflation that ultimately seed the temperature and matter-density variations observed today. 
These correlators can be expressed as linear combinations of more primitive objects called wavefunction coefficients, which exhibit fascinating mathematical structures, related to algebraic and positive geometry \cite{Arkani-Hamed:2017fdk, Glew:2025ypb, Capuano:2025ehm}, twisted (co)homology \cite{Arkani-Hamed:2023kig, De:2023xue, De:2024zic, Glew:2025ypb}, 
combinatorics \cite{Fevola:2024nzj, birkemeyer2026cosmologicalpolytopescanonicalforms}, 
and amplitudes \cite{He:2024olr, Baumann:2025qjx}.  

Auspiciously, wavefunction coefficients take a universal form in the class of theories under consideration. 
The flat-space wavefunction coefficient $\psi_\G^{(\mathrm{flat})}$ associated to a Feynman diagram $\G$ can be upgraded to a power-law FRW wavefunction coefficient $\psi_\G$ ($\vep \in\mathbb{C}\setminus\mathbb{Q}$)\footnote{When $\vep \in \mathbb{C}\setminus\mathbb{Q}$ the integral \eqref{eq:psiphys} is a twisted integral and we have good mathematical control over the resulting space of integrals. More specifically, easy access to a coaction \cite{Abreu:2019wzk}.}
by integrating against a kernel $u_\G$ called the twist:
\begin{align} \label{eq:psiphys}
    \psi_\G
    = \int_0^\infty u_\G \; \vphi_\G
    \,,\quad
    \vphi_\G
    = \psi^{(\mathrm{flat})}_\G
    (\mbf{X}{+}\mbf{x},\mbf{Y})\;
    \d^n\mbf{x}~.
\end{align}
They evaluate to hypergeometric functions that depend on the external energies $X_v$ that flow from each interaction vertex $v$ to the boundary, collectively denoted by $\mbf{X}$, and the energies $Y_e$ that are exchanged through each internal edge $e$, collectively denoted by $\mbf{Y}$. For example:
\begin{center}
\includegraphics[align=c, width=.32\textwidth]{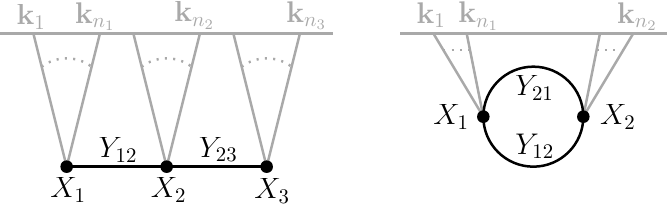}
\,.
\end{center}
With $\V_\G$ the set of vertices of $\G$, the twist
\begin{align}
    u_\G= \prod_{v\in \V_\G} x_v^{\alpha_v} \,,
\end{align}
is a multi-valued function whose exponents $\alpha_v \in \mathbb{C}\setminus\mathbb{Z}$ are determined by the underlying power-law cosmology (specified by $\vep$), the number of spatial dimensions $d$, and the valency $p_v$ of each vertex $v$ \cite{Arkani-Hamed:2023kig}.  
The simplest case occurs when $p_v=3$ and $d=3$, since then $\alpha_v =\vep$. 

The flat-space integrand $\vphi_\G$ is itself given by a simple combinatorial formula:
\begin{align} \label{eq:Psi}
    \vphi_\G
    = \d^{|\V_\G|} \mbf{x} \; \sum_{T_\G} \prod_{\tau \in T_\G} 
    \frac{ 1 }{ B_\tau } \, ,
\end{align}
where $T_\G$ denotes the set of (non-intersecting) complete tubings of $\G$. 
Each tube in a tubing is specified by a pair of sets $\tau = \{\mathcal{V}_\tau, \mathcal{E}_\tau\}$, where $\mathcal{V}_\tau$ is the set of vertices of $\G$ that are encircled by the tube $\tau$, while $\mathcal{E}_\tau$ is the set of edges that cross the tube $\tau$.
Given this data, we define the ``propagators''%
\footnote{%
    Technically, the $B_\tau$ are the square roots of propagators. However, we want the reader to approach the $B_\tau$ with the same intuition as if they were true propagators. 
}
\begin{align} \label{eq:singular_divisors}
    B_\tau = \sum_{v\in \mathcal{V}_\tau} x_v\,
    + \sum_{v\in \mathcal{V}_\tau}  X_v + \sum_{e\in \mathcal{E}_\tau} Y_e
    \,.
\end{align}
For more details on this combinatorial construction of $\vphi_\G$, see~\cite{Arkani-Hamed:2017fdk}.

In this letter, we show that the wavefunction of the universe---or, more directly, the wavefunction coefficients~\eqref{eq:psiphys}---satisfy a graphical coaction, by means of which their analytic structure can be easily understood. 
Our construction is similar in spirit to the diagrammatic coaction for Feynman integrals \cite{Abreu:2019wzk, Abreu:2017enx, Abreu:2017ptx, Abreu:2021vhb, Brown:2015fyf, Brown:2019jng}, but makes use of a slightly richer graphical language that keeps track of the direction of time (or energy flow). Roughly speaking, our graphical coaction, $\Delta$, takes the form 
\begin{align} 
    \Delta 
    \psi_\G
    &=\hspace{-1em} 
    \sum_{
        \text{decorations $\g$ of $\G$}
    }\hspace{-1em}
    \left(\substack{
        \text{rational}
        \\
        \text{function of } \alpha_v }
    \right)
    \big(\g \otimes \Csf_\g(\G)   \big) \, ,
\end{align}
where the sum is over all (acyclic) decorations $\g$ of the graph $\G$ (detailed in section \ref{sec:tubings_to_graphs}),
and $\Csf_\g(\G)$ represents the tubing of $\G$ that corresponds to $\g$. 
As we will show, each of the (decorated) graphs that appear on the right side of this formula have theory-independent integral interpretations; the integral associated with $\g$ encodes part of the differential equation satisfied by $\psi_\G$, while the graph 
$\Csf_\g(\G)$ is a cut integral that computes one of the (sequential) discontinuities of $\psi_\G$. In the $\alpha_v \to 0$ limit, our coaction reduces to the well-studied coaction on multiple polylogarithms (MPLs)~\cite{Chen,G91b,Goncharov:1998kja,Remiddi:1999ew,Borwein:1999js,Moch:2001zr,DelDuca:2009au,DelDuca:2010zg,Gonch2,Goncharov:2010jf,Brown:2011ik,Brown1102.1312,Frost:2023stm,Frost:2025lre}, which maps MPLs to their oft-utilized symbol when applied repeatedly~\cite{Goncharov:2010jf}. 
Note that, unlike~\cite{Abreu:2019wzk, Brown:2019jng}, we set all propagator powers to integers form the outset; thus, we work with physically relevant objects throughout.

\section{Physical Cuts and Acyclic Minors}
\label{sec:tubings_to_graphs}

For de Sitter universes involving only cubic interactions, wavefunction coefficients have an especially simple analytic structure, since (in three spatial dimensions) all $\alpha_v=\vep\to0$. 
Under these conditions, wavefunction coefficients evaluate to multiple polylogarithms (MPLs) of uniform transcendental weight $|\V_\G|$, with logarithmic branch points generated by the poles in the integrand where the propagators $B_\tau$ vanish. 
Therefore, computing a residue with respect to any of the polynomials $B_\tau$ isolates the discontinuity across a  logarithmic branch cut. 
More generally, trading the physical integration contour in~\eqref{eq:psiphys} for one that computes residues with respect to multiple propagators is equivalent to computing a sequence of discontinuities. 
These types of modified integrals---in which the integration contour wraps around the vanishing locus of a subset of the denominator factors---are often referred to as \emph{cut integrals}.

Only certain cut integrals are nonzero. This can already be seen in the structure of~\eqref{eq:Psi}, where pairs of propagators only appear in the same term if the corresponding tubes do not intersect. Additionally, the number of denominator factors in each term generally exceeds the number of integration variables $|\V_\G|$, implying we cannot compute residues with respect to all of them. Even so, the set of nonzero cuts of $\psi_\G$ are easily enumerated; they are in one-to-one correspondence with the \emph{acyclic minors} of $\G$~\cite{Glew:2025ypb}. Each acyclic minor corresponds to a decorated graph in which each edge in $\G$ is replaced by 
an \emph{oriented edge} 
($\adjustbox{valign=c}{\begin{tikzpicture}[scale=1]
    \coordinate (A) at (0,0);
    \coordinate (B) at (1/2,0);
    \coordinate (C) at (1,0);
    \coordinate (D) at (3/2,0);
    \draw[thick] (A) -- node{\ar} (B);
    \fill[black] (A) circle (2pt);
    \fill[black] (B) circle (2pt);
\end{tikzpicture}}$
or $\adjustbox{valign=c}{\begin{tikzpicture}[scale=1]
    \coordinate (A) at (0,0);
    \coordinate (B) at (1/2,0);
    \coordinate (C) at (1,0);
    \coordinate (D) at (3/2,0);
    \draw[thick] (A) -- node{\al} (B);
    \fill[black] (A) circle (2pt);
    \fill[black] (B) circle (2pt);
\end{tikzpicture}}$),
a \textit{pinched edge}  
($\adjustbox{valign=c}{\begin{tikzpicture}[scale=1]
    \coordinate (A) at (0,0);
    \coordinate (B) at (1/2,0);
    \coordinate (C) at (1,0);
    \coordinate (D) at (3/2,0);
    \draw[thick,double] (A) -- (B);
    \fill[black] (A) circle (2pt);
    \fill[black] (B) circle (2pt);
\end{tikzpicture}}$), 
or a \emph{disconnected/broken edge} ($\adjustbox{valign=c}{\begin{tikzpicture}[scale=1]
    \coordinate (A) at (0,0);
    \coordinate (B) at (1/2,0);
    \coordinate (C) at (1,0);
    \coordinate (D) at (3/2,0);
    \draw[thick,dotted] (A) -- (B);
    \fill[black] (A) circle (2pt);
    \fill[black] (B) circle (2pt);
\end{tikzpicture}}$), subject to the condition that there are no directed cycles in the resulting graph when all pinched edges are contracted. 
From a bulk physics point of view, we can think of the oriented edges as having a definite direction of time or energy flow, while the pinched and broken edges correspond to shrinking or eliminating the edge, respectively \cite{He:2024olr, Baumann:2025qjx}.

To each acyclic minor $\g$, we associate a set of \emph{cut tubings} $\Csf_\g$. These cut tubings $\csf\in\Csf_\g$ are maximal collections of non-crossing/compatible tubes such that 
    \vspace{-.5em}\begin{itemize}
    \item[(C1)] no $\tau \in \csf$ crosses a pinched edges of $\g$,
    \item[(C2)] no $\tau \in \csf$ encircles a broken edge of $\g$,
    \item[(C3)] any $\tau \in \csf$ crosses an edge $e$ only if  $e$ is a broken edge of $\g$ or $e$ is an oriented edge \includegraphics[align=c,scale=.3]{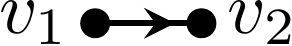} with $v_1 \in \V_\tau$ and $v_2 \not\in \V_\tau$. 
    \end{itemize}
For many acyclic minors, there is a unique tubing that satisfies these conditions; 
for example,
\begin{align}\begin{aligned}
    \Csf_{\includegraphics[align=c, scale=.3]{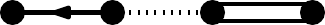}}
    = \left\{
        \includegraphics[align=c, scale=.3]{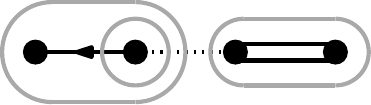}
    \right\}
    \,.
\end{aligned}\end{align}
However, some acyclic graphs have multiple tubings $\csf$, each of which satisfies (C1-C3), such as
\begin{align}
    \Csf_{\includegraphics[align=c, scale=.3]{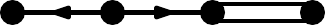}}
    &= \left\{
        \includegraphics[align=c, scale=.3]{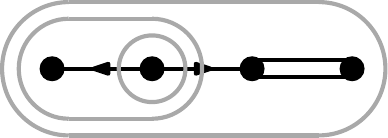}
        , 
        \includegraphics[align=c, scale=.3]{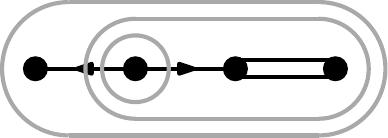}
    \right\}
    \,.
\end{align}
Importantly, even when multiple cut tubings belong to $\Csf_\g$, the geometry defined by the intersection $\bigcap_{\tau \in \csf} \{B_\tau{=}0\}$ is unique (different $\csf \in \Csf_\g$ just correspond to different ways of writing the same linear system of polynomial equations). This means that we can associate a unique ``physical'' residue operator with each acyclic minor $\g$:
\begin{align} \label{eq:degenRes}
    \res_\g 
    &= \sum_{\csf \in \Csf_\g}
        \sgn_\csf
        \res_\csf
        \,,
\end{align}
where the operator $\res_\csf$ computes a residue with respect to $B_\tau$ for all $\tau \in \csf$, and $\sgn_\csf \in \{1,-1\}$ is given in \eqref{eq:sgn}.  
Some examples include 
\begin{align}
        \res_{\includegraphics[align=c, scale=.3]{figs/lbs.pdf}}
    &= \res_{
        \includegraphics[align=c, scale=.2]{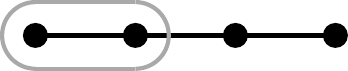}
        ,\includegraphics[align=c, scale=.2]{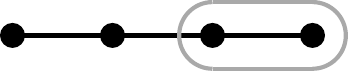}
        ,\includegraphics[align=c, scale=.2]{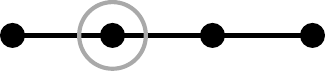}
    }
    \,,
    \\
    \res_{\includegraphics[align=c, scale=.3]{figs/lrs.pdf}}
    &= \res_{
        \includegraphics[align=c, scale=.2]{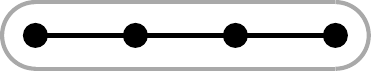},
        \includegraphics[align=c, scale=.2]{figs/4ChainB12.pdf},
        \includegraphics[align=c, scale=.2]{figs/4ChainB2.pdf}
    }
    \nn
    \\& \quad 
    - \res_{
        \includegraphics[align=c, scale=.2]{figs/4ChainB1234.pdf},
        \includegraphics[align=c, scale=.2]{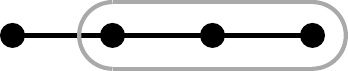},
        \includegraphics[align=c, scale=.2]{figs/4ChainB2.pdf}
    }
    \,.
\end{align}
Spelling out the above notation in terms of propagators, 
\be \label{eq:sequentialRes}
    \res_{\tau_1, \dots, \tau_m } :=
    \res_{\B_{\tau_2,\dots,\tau_m}} \circ \cdots \circ \res_{\B_{\tau_1}}\, ,
\ee
where $\B_{\tau_1,\dots,\tau_k} = \bigcap_{i=1}^k \{B_{\tau_i}=0\}$ is an intersection of hyperplanes. Importantly, we adopt a convention in which residues corresponding to \emph{larger} tubes (those that enclose more vertices) are computed before those of smaller tubes; geometrically, this amounts to a choice of the relative rate at which the corresponding hyperplanes are approached. Note that we also need to choose a convention for what region to integrate over once all residues in~\eqref{eq:degenRes} have been computed; we will have more to say about this in the next section.

The virtue of this acyclic minor construction is that it automatically takes into account all linear relations between partially-overlapping propagators, namely those generated by the relation
\begin{align}\label{eq:linRel}
    B_{\tau_1} + B_{\tau_2} 
    = B_{\tau_1 \cup \tau_2} + B_{\tau_1\cap\tau_2}
    \,.
\end{align} 
In other words, the residue operators $\res_\g$ associated with the acyclic minors of $\G$ span the space of operators that do not annihilate the physical wavefunction coefficient; all other residue operators either annihilate $\vphi_\G$, or are expressible as a linear combination of those specified by~\eqref{eq:degenRes}.

The same acyclic graph construction generalizes beyond de Sitter cosmologies, and to all polynomial interactions and spatial dimensions. For generic $\varepsilon \in \mathbb{C}\setminus\mathbb{Q}$, the coordinate hyperplanes $x_v = 0$ become singular surfaces of $u_\G$ in addition to playing the role of integration endpoints. While this changes the (co)homological underpinnings of the integral $\psi_\G$, the residue operators~\eqref{eq:degenRes} still span the space of cut integrals. The upshot is that acyclic graphs still provide a \emph{graphical representation} of the set of cuts---and by extension, the (sequential) discontinuities---of wavefunction coefficients for the full class of theories under consideration.%
\footnote{%
    The full space of integrands/contours---the (co)homology---is much larger than the physical subspace indexed by the acyclic minors. 
    While our graphical coaction applies only to this subspace, one can construct a non-graphical coaction on the full (co)homology from a generalization of \eqref{eq:schemCosCoaction}.
}

\section{From Cuts to the Coaction}

The full analytic structure of $\psi_\G$ is characterized not only by its sequential discontinuities, but also by its derivatives. 
In fact, more attention in the literature has been paid to the differential equations that describe wavefunction coefficients, and their formulation in terms of so-called kinematic flow~\cite{Arkani-Hamed:2023kig, Baumann:2025qjx, De:2023xue, De:2024zic, Glew:2025ypb, He:2024olr, Capuano:2025ehm}. 

Conveniently, a structure that combines knowledge of an integral's discontinuities and derivatives is known. 
The \emph{coaction}, best developed for MPLs~\cite{Goncharov:2010jf,Gonch2,Brown:2011ik,Brown1102.1312,Duhr:2012fh}, can be used to formally decompose twisted period integrals into (often simpler) building blocks whose analytic structure is already under good control. For wavefunction coefficients~\eqref{eq:psiphys} (and the integrals that encode their discontinuities and derivatives), this operation takes the form
\be \label{eq:schemCosCoaction}
    \Delta \int_\gamma u_\G\, \vphi
    =\sum_{\g,\h} C_{\g\h}^{-1}
    \tikz[baseline]{\node[basic,anchor=base] (tw) {$\displaystyle
        \overset{
            \partial_{\kin}
        }{\overbracket{
             \int_{\gamma} u_\G\, \vphi_\g 
    	}}
    $}}
    \otimes
    \tikz[baseline]{\node[basic,anchor=base] (tw) {$\displaystyle
        \overset{
            \disc_\kin
        }{\overbracket{
            \int_{\gamma_\h} u_\G\, \vphi
        }}
    $}}
    \,,
\ee
where the second entry is implicitly understood to be $\mathrm{mod}\;i\pi$, $\{\gamma_\g\}$ is the set of cut integration contours discussed in the last section, $\{\vphi_\h\}$ is a basis of differential forms that have poles at most where $\vphi_\G$ has poles, and $C_{\g\h}$ is the intersection matrix associated with these contours and forms~\cite{Abreu:2019wzk, Abreu:2017enx, Abreu:2017ptx, Abreu:2021vhb, Brown:2015fyf, Brown:2019jng}.%
\footnote{%
    We provide an explicit formula for the intersection numbers in \eqref{eq:Cmat}; more details will be provided in \cite{coactionLong}.
}
As indicated by the labeling of this formula, the first entry of the coaction encodes the kinematic derivatives of the original integral, while the second entry encodes its kinematic discontinuities. The upshot is that, if the coaction of $\psi_\G$ is known, its differential equation and discontinuities can be simply extracted. 

We have already seen that the discontinuities of the wavefunction coefficient $\psi_\G$ (or more specifically the set of cut contours $\{\gamma_\g\}$) can be given a graphical interpretation. 
After specifying these contours in more detail, we will now show that the remaining ingredients in~\eqref{eq:schemCosCoaction} can also be described graphically. 
This facilitates the construction of a graphical coaction for the wavefunction of the universe akin to the one for flat-space Feynman integrals~\cite{Abreu:2017enx,Abreu:2017mtm,Abreu:2021vhb}.

\subsection{A basis of contours from cuts}
\label{sec:cutContours}

Nontrivially, the set of residue operators~\eqref{eq:degenRes} can be used to build a basis of the homology on which $\psi_\G$ and its discontinuities are defined. Consequently, the physical integration contour $\smash{\gamma_\mathrm{phys} = \bigcap_{i\in \V_\G} \{x_i \ge 0\}}$ can be expressed as a linear combination of cut contours.
In order to fully specify these contours, however, we must identify the domain over which the remaining variables should be integrated, after all the residues in $\res_\g$ have been computed. To do so, we note that the twisted coordinate hyperplanes $\{x_i = 0\}$ that bound the original integration region always define a unique bounded region once a set of propagators have been set to zero.%
\footnote{%
    For twisted integrals ($\epsilon \in \mathbb{C}\setminus\mathbb{Q} \implies \alpha_v \in\mathbb{C}\setminus\mathbb{Z}$), the boundaries at $\{x_i=0\}$ are in the kernel of the twisted boundary operator.
    Therefore, twisted cycles are allowed to end on the coordinate hyperplanes. 
    One way to think about why contours are allowed to end on the twisted hyperplanes is that the exponents $\alpha_v$ regulate the integral; there exist a region in parameters where the integral converges and away from this region the integral is defined by analytic continuation.
} 
Therefore, for every acyclic minor $\g$ we define a cut contour $\gamma_\g$ via 
\be \label{eq:periodPairing1}
    \int_{\gamma_{\g}} u_\G\;\vphi
    &:= \int_{\Delta_\g} \res_{\g}[u_\G\;\vphi]
    \,,
\ee
where $\Delta_\g$ is the unique bounded region that has support on this cut.
Explicitly, 
\be 
    \Delta_\g {:=}  \Big\{\!
        (x_1,\dots,x_{|\V_\G|}) {\in} \mathbb{C}^{|\V_\G|}
        {\Big\vert} \substack{
            B_\tau {=} 0 \;\forall\; \tau\in\Csf_\g, 
            \\
            x_v \leq 0 \;\forall\; v \in\; \V^{\solidHalfEdge{.8}}_\g
        }
    \Big\}
    ,
\ee
where $\V^{\solidHalfEdge{.8}}_\g$ is the set of all vertices in $\g$ connected to at least one \emph{pinched} edge.

\subsection{A canonical basis of forms from cuts}
\label{sec:cutForms}

Having defined a basis of integration cycles $\{\gamma_\g\}$, we now identify a basis of differential forms $\{\vphi_\h\}$ that are dual to these cycles via integration. 
In fact, an appropriate basis of forms was constructed in~\cite{Glew:2025ypb}, which are also in one-to-one correspondence with the acyclic minors of $\G$. In particular, the form associated with the acyclic minor $\h$ is defined by having $\dlog$ singularities on all the propagators $B_\tau$ for $\tau \in \Csf$, as well as on the coordinate hyperplanes $x_v =0$ that bound the region $\Delta_\h$. One way to construct such a form is by finding a region that is bounded by all these hyperplanes, and computing its canonical form. In fact, there exist many such regions, and any choice would work. 

In practice, we find that the optimal choice is to set $\vphi_\g = \Omega[\check{\Gamma}_\g]$ equal to the canonical form of the (unbounded) region
\begin{align}
        \check{\Gamma}_\g 
    &:= \Big\{
        (x_1,{\dots},x_{|\V_\G|}) {\in} \mathbb{C}^{|\V_\G|}
        \Big\vert \substack{
            B_\tau \leq 0 \;\forall\; \tau\in\Csf_\g, 
            \\
            x_v \leq 0 \;\forall\; v \in\; \V^{\solidHalfEdge{.8}}_\g
            }
    \Big\} \,, 
\end{align}
since it results in a diagonal intersection matrix $\Csf_{\g\h}$. 
Further, note that since the maximal codimension boundary of $\check{\Gamma}_\g$ is $\Delta_\g$, the maximal cut of these differential forms is the canonical form of $\Delta_\g$
\be\label{eq:norm}
    \res_{\g}[\vphi_\g]
    = |\Csf_\g| \; \Omega[\Delta_\g]
    \,,
\ee
where $|\Csf_\g|$ is the cardinality of the set $\Csf_\g$.%
\footnote{%
    One can think of \eqref{eq:norm} as fixing the relative orientation between $\gamma_\g$ and $\check{\Gamma}_\g$. 
}

\subsection{Graphical FRW periods and their cuts}

Recalling \eqref{eq:periodPairing1}, the periods $P_{\g\h}$ of our basis are
\be
    P_{\g\h} := \int_{\gamma_\g} u_\G\; \vphi_\h 
        = \int_{\Delta_\g} \res_{\g}[u_\G\; \vphi_\h]
    \,.
\ee
Graphically, we denote the period $P_{\g\h}$ by superimposing the cut-tubing $\Csf_\g$ that represents the contour $\gamma_\g$ onto the acyclic minor $\h$ that represents the FRW form $\vphi_\h$. 
For example, in this graphical language, the ``diagonal'' periods of the two-site chain are
\begin{align} \label{eq:2chain_powerFns}
    \int_{\Delta_{\twoChaing}} \!\!\!\!\!\!\!\!
    \res_{\twoChaing}[u_{\G}\vphi_{\twoChaing}]
    &{=} \includegraphics[align=c, scale=.4]{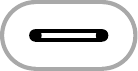}
    {=} \frac{\Gamma(\alpha_1)\Gamma(\alpha_2)}{\Gamma(\alpha_1{+}\alpha_2)} f_{\includegraphics[scale=.4]{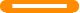}}^{\alpha_1+\alpha_2}
    ,
    \nn\\
    \int_{\Delta_{\twoChaingg}} \!\!\!\!\!\!\!\!
    \res_{\twoChaingg}[u_{\G} \vphi_{\twoChaingg}]
    &{=} \includegraphics[align=c, scale=.4]{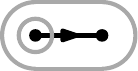}
    {=} f_{\includegraphics[scale=.4]{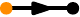}}^{\alpha_1}
    f_{\includegraphics[scale=.4]{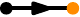}}^{\alpha_2}
    \,,
    \nn\\
    \int_{\Delta_{\twoChainggg}} \!\!\!\!\!\!\!\!
    \res_{\twoChainggg}[u_{\G} \vphi_{\twoChainggg}]
    &{=} \includegraphics[align=c, scale=.4]{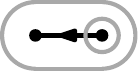}
    {=} f_{\includegraphics[scale=.4]{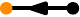}}^{\alpha_1}
    f_{\includegraphics[scale=.4]{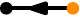}}^{\alpha_2}
    \,,
    \nn\\
    \int_{\Delta_{\twoChaingggg}} \!\!\!\!\!\!\!\!
    \res_{\twoChaingggg}[u_{\G} \vphi_{\twoChaingggg}]
    &{=} \includegraphics[align=c, scale=.4]{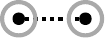}
    {=} f_{\includegraphics[scale=.4]{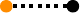}}^{\alpha_1}
    f_{\includegraphics[scale=.4]{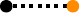}}^{\alpha_2}
    \,,
\end{align}
where 
\be
    f_{\includegraphics[scale=.4]{figs/2chain_Rs1}}
    {=} X_1 {+} X_2  
    ,\;\;\;
    f_{\includegraphics[scale=.4]{figs/2chain_Rr2}}
    {=} X_2 {-} Y_{12}
    ,\;\;\;
    f_{\includegraphics[scale=.4]{figs/2chain_Rl1}}
    {=} X_1 {-} Y_{12}
    ,
    \\
    f_{\includegraphics[scale=.4]{figs/2chain_Rr1}}
    {=} X_1 {+} Y_{12}
    {=} f_{\includegraphics[scale=.4]{figs/2chain_Rb1}}
    ,\;\;\;
    f_{\includegraphics[scale=.4]{figs/2chain_Rl2}}
    {=} X_2 {+} Y_{12}
    {=} f_{\includegraphics[scale=.4]{figs/2chain_Rb2}}
    ,
\ee
are the symbol letters that appear in the differential equations (see \cite{Glew:2025ypb} for more on this notation).
The remaining non-zero periods for the two-site chain evaluate to hypergeometric functions. 
In the kinematic region where $X_1, X_2 > Y_{12} > 0$,  
\begin{align}\label{eq:2chain_2F1s}\begin{aligned}
    \frac{
        \includegraphics[align=c, scale=.4]{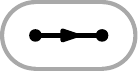}
    }{
        \includegraphics[align=c, scale=.4]{figs/2chainP11.pdf}
    }
    {=}
    c\;
    {}_2F_1\!\!\left(
        1,
        1{+}\alpha_1;
        2{+}\alpha_1{+}\alpha_2;
        \frac{
            f_{\includegraphics[scale=.4]{figs/2chain_Rs1}}
        }{
            f_{\includegraphics[scale=.4]{figs/2chain_Rr1}}
        }
    \right)
    \!\!
    \frac{
        f_{\includegraphics[scale=.4]{figs/2chain_Rs1}}
    }{
        f_{\includegraphics[scale=.4]{figs/2chain_Rr1}}
    }
    ,
    \\
    \frac{
        \includegraphics[align=c, scale=.4]{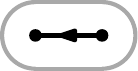}
    }{
        \includegraphics[align=c, scale=.4]{figs/2chainP11.pdf}
    }
    {=}
    c\;
    {}_2F_1\!\!\left(
        1,
        1{+}\alpha_2;
        2{+}\alpha_1{+}\alpha_2;
        \frac{
            f_{\includegraphics[scale=.4]{figs/2chain_Rs1}}
        }{
            f_{\includegraphics[scale=.4]{figs/2chain_Rl2}}
        }
    \right)
    \!\!
    \frac{
        f_{\includegraphics[scale=.4]{figs/2chain_Rs1}}
    }{
        f_{\includegraphics[scale=.4]{figs/2chain_Rl2}}
    }
    ,
\end{aligned}\end{align}
for 
$c = -\alpha_1\alpha_2 \Gamma(\alpha_1{+}\alpha_2) / \Gamma(2{+}\alpha_1{+}\alpha_2)$.

Note that periods with tubes that cross a pinched edge, or an oriented edge that points the wrong way, vanish:
\be
    0&=\includegraphics[align=c,scale=.4]{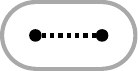}
    = \includegraphics[align=c,scale=.4]{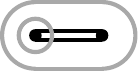}
    = \includegraphics[align=c,scale=.4]{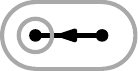}
    = \includegraphics[align=c,scale=.4]{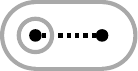}
    = \includegraphics[align=c,scale=.4]{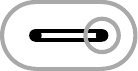}
    \\&\quad
    = \includegraphics[align=c,scale=.4]{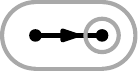}
    = \includegraphics[align=c,scale=.4]{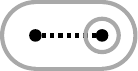}
    = \includegraphics[align=c,scale=.4]{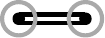}
    = \includegraphics[align=c,scale=.4]{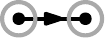}
    = \includegraphics[align=c,scale=.4]{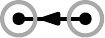}
    \,.
\ee
This is the graphical translation of the fact that 
\be \label{eq:vanishingCuts}
    \res_\mathfrak{\g}[\vphi_\h] = 0 
    \;\forall\; 
    \g \notin \mathrm{pinch}(\h)
    \,,
\ee
where $\mathrm{pinch}(\g)$ is the set of all acyclic minors obtained from $\g$ by turning any number of oriented edges into pinched edges (including no edges). 
In other words, $P_{\g\h}\neq0$ if and only if $\g$ is a pinch of $\h$.

\section{The graphical coaction}
\label{sec:graphicalCoaction}

Rephrasing the coaction \eqref{eq:schemCosCoaction} in terms of the contours and forms defined in the last section, we get 
\be \label{eq:coactionMaster}
    \Delta 
    \int_{\gamma_\g} 
    u_\G\, \vphi_\mathfrak{h}
    &= 
    \sum_{
        \mathfrak{f} 
    }
    C_{\mathfrak{f}\mathfrak{f}}^{-1}
    \int_{\gamma_\g} 
    u_\G\, \vphi_\mathfrak{f}
    \otimes
    \int_{\gamma_\mathfrak{f}} 
    u_\G\, \vphi_\h
    \,,
\ee
where the right entry is again understood modulo $i\pi$, while~\cite{coactionLong, Glew:2025ypb}
\be\label{eq:Cmat}
    C_{\g\h} 
    &:= \delta_{\g\h}\;  
        |\Csf_\g| \; 
        \prod_{\rsf \in \mathrm{Sub}^{\includegraphics[scale=.3]{figs/2chain_gs}}_\g} 
        \frac{
            \sum_{v\in\V_\rsf} \alpha_v
        }{
            \prod_{v\in\V_\rsf} \alpha_v
        }
    \,,
\ee
and $\delta_{\g\h}$ is the Kronecker delta symbol. 
Here, $C_{\g\g}$ are intersection numbers associated to our basis choice and  $\mathrm{Sub}^{\twoChaing}_\g$ is the set of all connected subgraphs of $\g$ with pinched edges. 
Due to \eqref{eq:vanishingCuts} only the terms in the sum with $\g \in \mathrm{pinch}(\mathfrak{f})$ and $\mathfrak{f} \in \mathrm{pinch}(\h)$ are non-vanishing.

\subsection{The two-site chain graph}

The coaction of any power function (such as seen in \eqref{eq:2chain_powerFns}) is trivial; both their derivative and discontinuity are proportional to themselves. 
For example, 
\be
    \Delta 
    \left[\includegraphics[align=c, scale=.4]{figs/2chainP11.pdf}\right]
    &= 
    \frac{\alpha_1\alpha_2}{\alpha_1{+}\alpha_2} 
    \includegraphics[align=c, scale=.4]{figs/2chainP11.pdf}
    \otimes 
    \includegraphics[align=c, scale=.4]{figs/2chainP11.pdf}
    \,,
    \\
    \Delta 
    \left[\includegraphics[align=c, scale=.4]{figs/2chainP22.pdf}\right]
    &= 
    \includegraphics[align=c, scale=.4]{figs/2chainP22.pdf}
    \otimes 
    \includegraphics[align=c, scale=.4]{figs/2chainP22.pdf}
    \,.
\ee
Conversely, the hypergeometric functions \eqref{eq:2chain_2F1s} enjoy a more interesting coaction. 
For example,
\be \label{eq:Delta2F1}
    \Delta 
    \includegraphics[align=c, scale=.4]{figs/2chainP12.pdf}
    {=} \frac{\alpha_1\alpha_2}{\alpha_1{+}\alpha_2} 
    \includegraphics[align=c, scale=.4]{figs/2chainP11.pdf}
    {\otimes}
    \includegraphics[align=c, scale=.4]{figs/2chainP12.pdf}
    +
    \includegraphics[align=c, scale=.4]{figs/2chainP12.pdf}
    {\otimes} 
    \includegraphics[align=c, scale=.4]{figs/2chainP22.pdf}
    \,.
\ee
Indeed, the derivative/discontinuity of a ${}_2F_1$ can be proportional to either itself or to a power function.
For example, extracting the weight-one component of the second entry in \eqref{eq:Delta2F1} yields the differential equation satisfied by \includegraphics[align=c, scale=.3]{figs/2chainP12.pdf}:
\vspace{-.5em}
\be
    \d[\includegraphics[align=c, scale=.4]{figs/2chainP12.pdf}]
    &{=} \includegraphics[align=c, scale=.4]{figs/2chainP11.pdf}\,
        \overset{
            \frac{\alpha_1\alpha_2}{\alpha_1{+}\alpha_2} 
            \d W_1[\includegraphics[align=c, scale=.3]{figs/2chainP12.pdf}]
        }{\overbracket{
        \left( 
            \frac{\alpha_1\alpha_2}{\alpha_1{+}\alpha_2}
            \dlog\frac{
                f_{\includegraphics[scale=.4]{figs/2chain_Rr2}}
            }{
                f_{\includegraphics[scale=.4]{figs/2chain_Rr1}}
            }
        \right) 
        }}
    \\&
    \quad {+} \includegraphics[align=c, scale=.4]{figs/2chainP12.pdf}\,
        \underset{
             \d W_1[\includegraphics[align=c, scale=.3]{figs/2chainP22.pdf}]
        }{\underbracket{
        \left( 
            \alpha_1 
            \dlog f_{\includegraphics[scale=.4]{figs/2chain_Rr2}}
            {+} \alpha_2
            \dlog f_{\includegraphics[scale=.4]{figs/2chain_Rr1}}
        \right)
        }}
    \,,
\ee
This agrees with \cite{Glew:2025ypb}. 
Similarly, any discontinuity of \includegraphics[align=c, scale=.3]{figs/2chainP12.pdf} can be obtained by extracting the weight-one part of the first entry:
\vspace{-.5em}
\be
    \disc[\includegraphics[align=c, scale=.4]{figs/2chainP12.pdf}]
    &{=}
        \overset{
             \frac{\alpha_1\alpha_2}{\alpha_1{+}\alpha_2} 
            \disc W_1[\includegraphics[align=c, scale=.3]{figs/2chainP11.pdf}]
        }{\overbracket{
            (\alpha_1 {+} \alpha_2) 
            \left( 
                \disc \log f_{\includegraphics[scale=.4]{figs/2chain_Rs1}}
            \right) 
        }}\,
        \includegraphics[align=c, scale=.4]{figs/2chainP12.pdf}
    \\&
    \quad {+} 
        \underset{
            \disc W_1[\includegraphics[align=c, scale=.3]{figs/2chainP12.pdf}]
        }{\underbracket{
        \left( 
            \disc \log \frac{
                f_{\includegraphics[scale=.4]{figs/2chain_Rr2}}
            }{
                f_{\includegraphics[scale=.4]{figs/2chain_Rr1}}
            }
        \right)
        }}\,
        \includegraphics[align=c, scale=.4]{figs/2chainP22.pdf}
    \,.
\ee
Here, $\disc$ is a stand-in for a specific choice of discontinuity, for instance $\disc_{f_{\includegraphics[scale=.4]{figs/2chain_Rs1}}=0}$. 

Note that, to make use of~\eqref{eq:coactionMaster} for the full wavefunction coefficient $\psi_\G$, we should first decompose the physical contour $\gamma_\mathrm{phys}$ and physical differential form $\vphi_\G$ into our chosen bases $\{\gamma_\g\}$ and $\{\vphi_\g\}$. This decomposition is provided for the two-site chain in appendix \ref{app:2chainDets}.%
\footnote{The decomposition of any $\vphi_\G$ into our basis is trivial---a closed form formula exists \cite{Glew:2025arc,Glew:2025ypb,coactionLong}. On the other hand, the decomposition of $\gamma_\mathrm{phys}$ straightforward but tedious. 
} 
From this, the coaction on the physical integral is deduced.

\subsection{A one-loop example}

To illustrate the structure of the graphical coaction beyond tree-level, consider the following one-loop example 
\allowdisplaybreaks
\begin{align}
    &\Delta 
    \;\includegraphics[align=c, scale=.4]{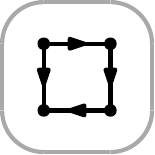}\;
    {=} 
    \;\includegraphics[align=c, scale=.4]{figs/C1234Box.pdf}\;
    {\otimes}
    \;\includegraphics[align=c, scale=.4]{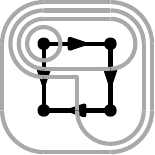}\;
    {+} \alpha_{12}
    \;\includegraphics[align=c, scale=.4]{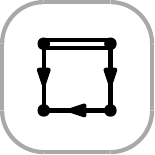}\;
    {\otimes}
    \;\includegraphics[align=c, scale=.4]{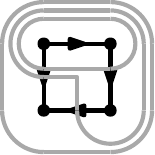}\;
    \nn
    \\& 
    {+} \alpha_{23}
    \;\includegraphics[align=c, scale=.4]{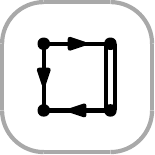}\;
    {\otimes}
    \;\includegraphics[align=c, scale=.4]{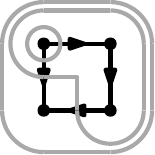}\;
    {+} \alpha_{34}
    \;\includegraphics[align=c, scale=.4]{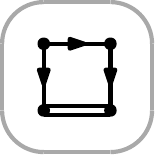}\;
    {\otimes}
    \;\includegraphics[align=c, scale=.4]{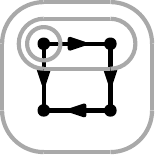}\;
    \\& 
    {+} \alpha_{123}
    \;\includegraphics[align=c, scale=.4]{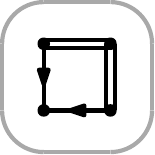}\;
    {\otimes}
    \;\includegraphics[align=c, scale=.4]{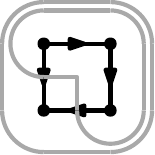}\;
    {+} \alpha_{234}
    \;\includegraphics[align=c, scale=.4]{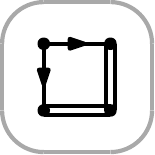}\;
    {\otimes}
    \;\includegraphics[align=c, scale=.4]{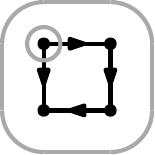}\;
    \nn
    \\& 
    {+} \alpha_{12}\alpha_{34}
    \;\includegraphics[align=c, scale=.4]{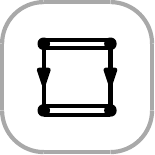}\;
    {\otimes}
    \;\includegraphics[align=c, scale=.4]{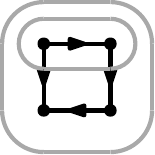}\;
    {+} \alpha_{1234}
    \;\includegraphics[align=c, scale=.4]{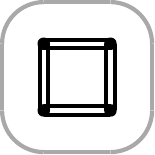}\;
    {\otimes}
    \;\includegraphics[align=c, scale=.4]{figs/C1234Box.pdf}\;
    ,
    \nn
\end{align}
where $\alpha_{ij\cdots}:=\frac{\alpha_i\alpha_j\cdots}{\alpha_i+\alpha_j+\cdots}$.
Note that with loop graphs, the acyclic condition actually matters. 
The decorated minors 
\be
    {}&\bigg\{
    \includegraphics[align=c, scale=.4]{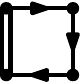}, 
    \includegraphics[align=c, scale=.4]{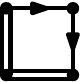},
    \includegraphics[align=c, scale=.4]{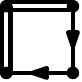},
    \includegraphics[align=c, scale=.4]{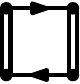},
    \includegraphics[align=c, scale=.4]{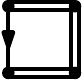},
    \includegraphics[align=c, scale=.4]{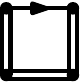},
    \includegraphics[align=c, scale=.4]{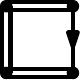},
    \includegraphics[align=c, scale=.4]{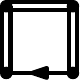}
    \bigg\}
    \,,
\ee
do not appear because they are cyclic once the pinched edges are contracted.

\subsection{Further Examples and Cross-Checks}

A more complete derivation of~\eqref{eq:coactionMaster}, as well as further examples, will be provided in a forthcoming publication~\cite{coactionLong}. In particular, we have checked (in a number of nontrivial examples) that our coaction formula commutes with the well-known MPL coaction, when expanded in $\alpha_v$. In~\cite{coactionLong}, we also derive an explicit formula for the weight-one part of any period $P_{\g\h}$, allowing us to show that the kinematic flow construction follows directly from~\eqref{eq:coactionMaster}.

\section{Conclusion}

In this letter, we have formulated a graphical coaction for the wavefunction of the universe in theories of conformally coupled scalars in power-law FRW cosmologies, valid at all particle multiplicities and arbitrary loop orders. By leveraging the twisted (co)homology of the associated integral families, the coaction decomposes wavefunction coefficients into tensor products of simpler FRW periods, each represented by a decoration of the original Feynman diagram. This yields a unified, graphical language that encapsulates both the differential equations and the sequential discontinuities of cosmological correlators. Our construction is analogous to (and may provide hints for studying) the diagrammatic coaction for flat-space Feynman integrals, but involves a richer graphical language that incorporates the direction of time. 

Our construction has been implemented in a user-friendly web application at \webappAlt\!\!.
This application can handle user defined examples and generates explicit expressions for the graphical building blocks.

Several natural extensions remain. It would be interesting to connect this construction more explicitly to the graphical coaction for flat-space Feynman integrals in the $\vep\to-1$ limit, and to explore whether the coaction structure persists for theories with spinning fields, non-power-law scale factors or non-conformally coupled scalars. We leave these fascinating directions for future work.


\acknowledgments
The authors would like to thank D.~Baumann, R.~Britto, and C.~Dupont for stimulating discussions. 
We also thank C.~Duhr, whose original suggestion led to this work.
AJM is supported by the Royal Society grant URF{\textbackslash}R1{\textbackslash}221233, and additionally acknowledges support from the European Research Council (ERC) under the European Union’s Horizon Europe research and innovation program grant agreement 101163627 (ERC Starting Grant
“AmpBoot”).
AP is supported by the European Union (ERC, UNIVERSE PLUS, 101118787). 
LR is supported by the Royal Society via a Newton International Fellowship. 

\appendix 

\section{Definition of $\sgn_\csf$}

In this appendix, we define $\sgn_\csf$.
While easy to compute, the definition requires the introduction of new ideas that are not needed for the main text.

Let  $\sigma$ be the permutation that orders the tubes $\tau\in\csf$ according to size. 
Also let $\Rsf_\g = \mathrm{Sub}^{\twoChaing}_\g \cup \mathrm{Sub}^{\bullet}_\g$ 
where $\mathrm{Sub}^{\twoChaing}_\g$ is the collection of all connected subgraphs of $\g$ with only \emph{pinched} edges 
and $\mathrm{Sub}^{\bullet}_\g$ is the set of all 1-vertex subgraphs attached to no pinched edges. 
Then, with $\rsf_i \in \Rsf_\g$, 
\begin{align} \label{eq:sgn}
    \sgn_\csf &:= 
    \frac{
        \mathrm{sig}\left(
            \la \bar{\rsf}_1 \ra_\csf, 
            \dots, 
            \la \bar{\rsf}_{|\Rsf_\g|} \ra_\csf
        \right) 
    }{
        \mathrm{sig}\left(
            \sigma\left( \la \bar{\rsf}_{1} \ra_\csf \right), 
            \dots, 
            \sigma\left( \la \bar{\rsf}_{|\Rsf_\g|} \ra_\csf \right)
        \right) 
    }
    \,,
\end{align}
where $\bar{\rsf}_i := \min \V_\rsf$ is minimal vertex of the subgraph $\rsf_i$.
 We also define $\la v \ra_\csf$ as the smallest tube $\tau\in\csf$ that contains the subgraph $\rsf$ that also contains the vertex $v$:%
\be
    \la v \ra_\csf 
    &:= \tau \in \csf 
    \text{ such that }
    v \in \V_\rsf 
    \\&\qquad
    \text{ and } 
    |\V_\rsf| > |\V_{\rsf^\prime}| 
    \;\forall\; \rsf^\prime \in \tau 
    \text{ with } \rsf^\prime \neq \rsf
    \,.
\ee
Note that the map $\la v \ra_\csf : \V_\g \to \csf$ is many-to-one whenever the associated minor $\g$ contains pinched edges.

\section{The two-site chain wavefunction coefficient in terms of FRW periods}
\label{app:2chainDets}

The physical wavefunction coefficient $\psi_{\G}$ can be written in terms of the FRW periods \eqref{eq:2chain_powerFns} and \eqref{eq:2chain_2F1s}. 
First, one decomposes the physical differential form into the $\vphi_\g$ using partial fractions
\be \label{eq:2chainPartialFraction}
    \vphi_{\mathrm{2-chain}}
    = \vphi_{\twoChaingg} + \vphi_{\twoChainggg} 
    - \vphi_{\twoChaingggg}
    \,.
\ee
The general partial fraction decomposition has a closed form expression (see \cite{Glew:2025ypb}). 
We can also decompose the physical contour into our basis of cycles. 
While we do not have a general formula, it can be done case-by-case when needed
\begin{align} \label{eq:2chainContourDecomp}
    \gamma_\mathrm{phys}^{\mathrm{2-chain}}
    &= \frac{
        -\pi e^{- \pi i (\alpha_1{+}\alpha_2)} 
        \gamma_{\twoChaing}
    }{
        \sin\big(\pi (\alpha_1{+}\alpha_2) \big)
    }
    \nn\\&
    + \frac{
            \pi^2 e^{- \pi i \alpha_2} 
            \gamma_{\twoChaingg}
        }{
            \sin\big(\pi \alpha_1 \big) 
            \sin\big(\pi (\alpha_1{+}\alpha_2) \big) 
        }
    \nn\\& 
    + \frac{
        \pi^2 e^{- \pi i \alpha_1}  
        \gamma_{\twoChainggg}
    }{
        \sin(\pi \alpha_2) 
        \sin\big(\pi (\alpha_1{+}\alpha_2) \big)
    }
    \nn\\&
    + \frac{
        \pi^2 e^{- \pi i (\alpha_1{+}\alpha_2)} 
        \gamma_{\twoChaingggg}
    }{
        \sin(\pi \alpha_1) \sin(\pi \alpha_2)
    }
\end{align}
Combining \eqref{eq:2chainContourDecomp} with \eqref{eq:2chainPartialFraction}, yields 
\begin{align}
    \psi_{\mathrm{2-chain}} 
    &= \frac{
            -\pi e^{- \pi i (\alpha_1{+}\alpha_2)} 
        }{
            \sin\big(\pi (\alpha_1{+}\alpha_2) \big)
        }
        \left[
            \includegraphics[align=c, scale=.5]{figs/2chainP12.pdf}
            {+}\includegraphics[align=c, scale=.5]{figs/2chainP13.pdf}
        \right]
    \nn\\&
    + \frac{
            \pi^2 e^{- \pi i \alpha_2} 
        }{
            \sin\big(\pi \alpha_1 \big) 
            \sin\big(\pi (\alpha_1{+}\alpha_2) \big) 
        } 
        \includegraphics[align=c, scale=.5]{figs/2chainP22.pdf}
    \nn\\&
    + \frac{
            \pi^2 e^{- \pi i \alpha_1}  
        }{
            \sin(\pi \alpha_2) 
            \sin\big(\pi (\alpha_1{+}\alpha_2) \big)
        } 
        \includegraphics[align=c, scale=.5]{figs/2chainP33.pdf}
    \nn\\&
    - \frac{
        \pi^2 e^{- \pi i (\alpha_1{+}\alpha_2)} 
    }{
        \sin(\pi \alpha_1) \sin(\pi \alpha_2) 
    } 
    \includegraphics[align=c, scale=.5]{figs/2chainP44.pdf}
    \,.
\end{align}
\hspace{20em}

\bibliographystyle{apsrev4-1.bst}
\bibliography{reference.bib}
\end{document}